\documentclass[prl,reprint,a4paper,superscriptaddress,aps]{revtex4-1}

\usepackage{graphicx}
\graphicspath{{/home/user/fig/publi/muon/}}
\usepackage{dcolumn}
\usepackage{bm}

\begin{document}


\title{Core-shell nanostructure in a Ge$_{0.9}$Mn$_{0.1}$ film from structural and magnetic measurements 
}

\author{P.~Dalmas de R\'eotier}
\affiliation{Univ. Grenoble Alpes, INAC-SPSMS, F-38000 Grenoble, France}
\affiliation{CEA, INAC-SPSMS, F-38000 Grenoble, France}
\author{E.~Prestat}
\affiliation{Univ. Grenoble Alpes, INAC-SP2M, F-38000 Grenoble, France}
\affiliation{CEA, INAC-SP2M, F-38000 Grenoble, France}
\author{P.~Bayle-Guillemaud}
\affiliation{Univ. Grenoble Alpes, INAC-SP2M, F-38000 Grenoble, France}
\affiliation{CEA, INAC-SP2M, F-38000 Grenoble, France}
\author{A.~Barski}
\affiliation{Univ. Grenoble Alpes, INAC-SP2M, F-38000 Grenoble, France}
\affiliation{CEA, INAC-SP2M, F-38000 Grenoble, France}
\author{A.~Marty}
\affiliation{Univ. Grenoble Alpes, INAC-SP2M, F-38000 Grenoble, France}
\affiliation{CEA, INAC-SP2M, F-38000 Grenoble, France}
\author{M.~Jamet}
\affiliation{Univ. Grenoble Alpes, INAC-SP2M, F-38000 Grenoble, France}
\affiliation{CEA, INAC-SP2M, F-38000 Grenoble, France}
\author{A.~Suter}
\affiliation{Laboratory for Muon-Spin Spectroscopy, Paul Scherrer Institute, 
CH-5232 Villigen-PSI, Switzerland}
\author{T.~Prokscha}
\affiliation{Laboratory for Muon-Spin Spectroscopy, Paul Scherrer Institute, 
CH-5232 Villigen-PSI, Switzerland}
\author{Z.~Salman}
\affiliation{Laboratory for Muon-Spin Spectroscopy, Paul Scherrer Institute, 
CH-5232 Villigen-PSI, Switzerland}
\author{E.~Morenzoni}
\affiliation{Laboratory for Muon-Spin Spectroscopy, Paul Scherrer Institute, 
CH-5232 Villigen-PSI, Switzerland}
\author{A. Yaouanc}
\affiliation{Univ. Grenoble Alpes, INAC-SPSMS, F-38000 Grenoble, France}
\affiliation{CEA, INAC-SPSMS, F-38000 Grenoble, France}

\date{\today}

\begin{abstract}
We have characterized a film of Ge$_{0.9}$Mn$_{0.1}$ forming self-organized nanocolumns perpendicular to the Ge substrate with high 
resolution scanning transmission electron microscopy combined with electron energy loss spectroscopy, 
and bulk magnetization and positive muon spin rotation and relaxation ($\mu$SR) measurements. The Mn-rich nanocolumns form a triangular lattice with no detectable Mn atoms in the matrix. They consist of cores surrounded by shells. The combined analysis of bulk magnetization and $\mu$SR data enables us to characterize the electronic and magnetic properties of both the cores and shells. The discovered phase separation of the columns between a core and a shell is probably relevant for other transition-metal doped semiconductors.

\end{abstract}


\maketitle

Spin-polarized carriers in spintronic applications may be conveyed by conventional metallic 
ferromagnets or by so-called ferromagnetic semiconductors. Until now semiconductor spintronics 
has mainly been based on diluted magnetic semiconductors \cite{Timm03}. Another route consists in the use of films with a well-defined pattern of transition metal-rich nanostructures with a relatively high-temperature ferromagnetic phase \cite{Bougeard06,Kuroda07,Bonanni08}. Nanocolumns as in Ge$_{1-x}$Mn$_x$ films \cite{Jamet06,Li07} are one of these structures.

The magnetic properties of these films have so far been investigated by 
x-ray magnetic circular dichroism \cite{Tardif10a,Tardif14} and electron spin 
resonance \cite{Jain10,Jain11}, but the most conclusive results stem from superconducting quantum 
interference device (SQUID) magnetometry \cite{Devillers07,Yu10}. Here we resolve the structure and 
characterize the magnetism of the nanocolumns of a Ge$_{0.9}$Mn$_{0.1}$ film at the atomic scale using  
high resolution scanning transmission electron microscopy (STEM) supplemented with electron energy 
loss spectroscopy (EELS) analysis, and SQUID and muon spin rotation and relaxation ($\mu$SR) measurements. We find the 
film to consist of nanocolumns embedded in a Ge matrix with less than 0.05~at.\,\% of Mn. Each nanocolumn 
is made of a core surrounded by a shell with notable different magnetic properties. We suggest that this three-fold structural and magnetic phase separation is generic to impurity doped semiconductors with spinodal decomposition.

Our 80 nm-thick Ge$_{0.9}$Mn$_{0.1}$ film has been grown by molecular beam epitaxy at low temperature. Ge 
and Mn atoms have been co-evaporated at 100$^\circ$C on a Ge(100) substrate using standard Knudsen cells. 
Due to the very low solubility of Mn in Ge, a spontaneous two-dimensional spinodal decomposition takes place 
within the film at the early stage of the growth \cite{Fukushima06}. The following layer-by-layer growth 
leads to the formation of Mn-rich nanocolumns spanning the whole film thickness \cite{Jamet06}. Their 
average diameter is $d_{\rm nc} = 4$~nm --- see the images in the supplemental material \cite{SM}.
The nanocolumns form a slightly disordered triangular lattice with an average lattice 
parameter $a^{\rm lp}_{\rm nc}$ = 10~nm.

The STEM-EELS analysis was performed at room temperature with an {\aa}ngstr\"om-sized electron probe 
\cite{SM}. The $\mu$SR spectra were recorded from 300 down to 5~K at the 
low energy muon (LEM) spectrometer \cite{Morenzoni04,Prokscha08,SM} of the  Swiss Muon Source (S$\mu$S, Paul Scherrer Institute, Switzerland) with a 9~cm$^2$ area sample. The spectra were taken either in zero or in a finite external field ${\bf B}_{\rm ext}$ applied perpendicular to the film substrate.

\begin{figure}[tb]
\begin{center}
\includegraphics[width=0.48\textwidth]{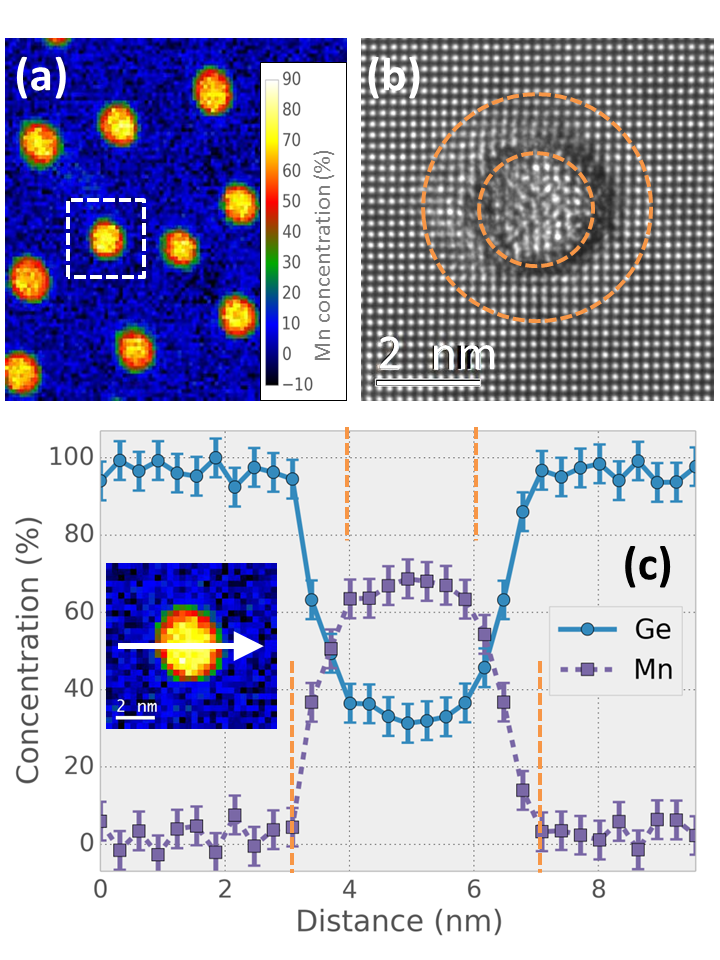}
\end{center}
\caption{(Color online) Characterization of our Ge$_{0.9}$Mn$_{0.1}$ film. (a) Mn 
concentration map obtained by EELS. (b) STEM plane view. (c) Ge and Mn concentration profiles across a 
single nanocolumn. In (b) and (c) the orange dashed circles and vertical dashed lines specify the 
nanocolumn core and shell positions.}
\label{fig_structure}
\end{figure}

Results of STEM-EELS measurements are presented in Fig.~\ref{fig_structure}.
As shown in Fig.~\ref{fig_structure}b, the nanocolumns exhibit a rather complex inner crystal structure and 
are surrounded by a Ge matrix with huge tensile strain up to $1.0 \, (3)$\%. The Mn content 
within the nanocolumns is not uniform. As shown in Fig.~\ref{fig_structure}c, they exhibit a core-shell 
structure with a $d_{\rm c} = 2$~nm core diameter containing up to 70~at.\,\% of Mn surrounded with a 
$r_{\rm s} = 1$~nm thick shell in which the Mn content decreases roughly linearly from 70 to less 
than 0.05~at.\,\%, i.e.\ the detection limit, in the matrix. The average Mn content of $33$~at.~\%
over the total area 
agrees with a previous estimate \cite{Mouton12}. Note the nanocolumn core high atomic density: 
$82 \, (8)$~atoms/nm$^3$ to be compared to 44 atoms/nm$^3$ in pure Ge. A close packed metallic 
structure is therefore formed in the nanocolumns. It is tempting to assign it to the most stable Ge-Mn alloy 
of formula Ge$_3$Mn$_5$ since its Mn atomic fraction of 5/8 is close to the value measured in 
the nanocolumns core \cite{Rovezzi08}. However, x-ray diffraction have 
definitely excluded the presence of the hexagonal Ge$_3$Mn$_5$ phase within the film \cite{Tardif10b}. 
Magnetization measurements discussed later on also support this result since the magnetic moment per Mn in the 
core is much less than 2.6 $\mu_{\rm B}$/Mn in pure Ge$_3$Mn$_5$. 
The Mn atomic density varies from $57 \, (5)$~atoms/nm$^3$  
in the nanocolumns core down to approximately zero in the Ge matrix. Assuming a linear Mn concentration 
profile in the shell, we find that 3/7 (4/7) of Mn atoms are located in the nanocolumns core (shell). 

\begin{figure}[tb]
\begin{center}
\includegraphics[width=0.48\textwidth]{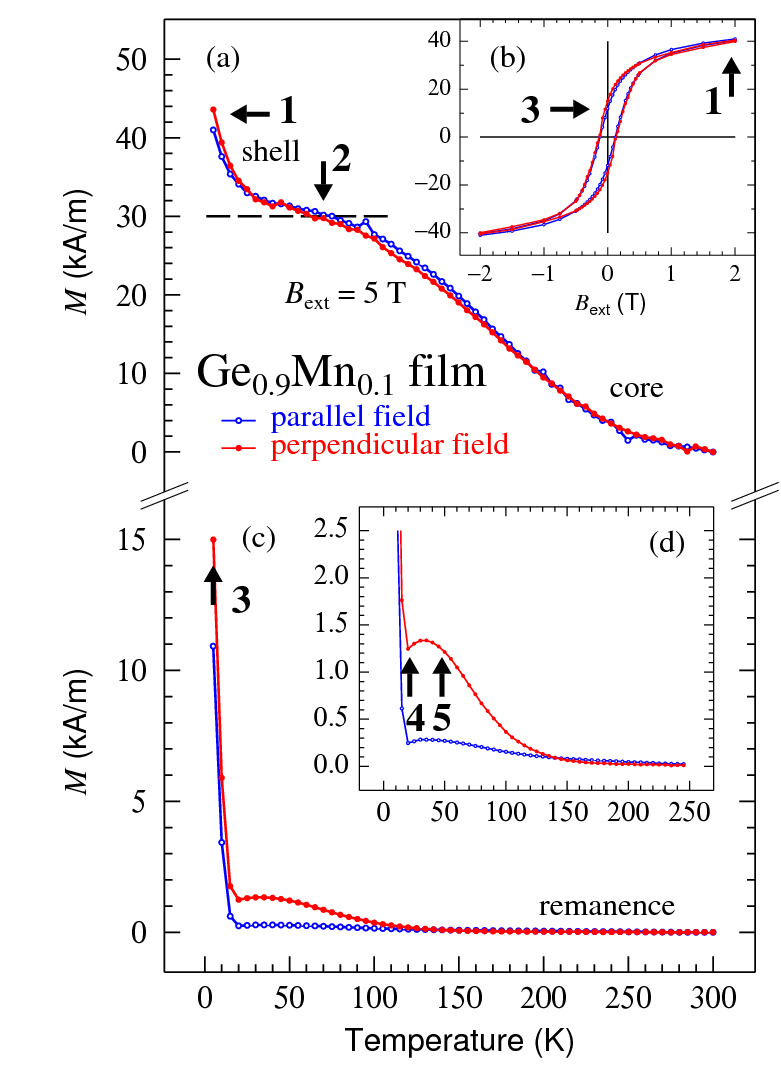}
\end{center}
\caption{(Color online) Bulk magnetization measurements for our Ge$_{0.9}$Mn$_{0.1}$ film normalized
to its volume. ${\bf B}_{\rm ext}$ is either parallel or perpendicular to the film substrate. (a) 
Saturation magnetization under 5~T as a function of temperature. 
(b) Magnetization curves recorded at 5~K. (c) Remanent magnetization as a function of temperature 
after field cooling under 5~T. For the sake of clarity the magnetization between 5 and 250~K has been 
enlarged in (d), emphasizing the unexpected uprise of the magnetization above 20~K. The different magnetic configurations denoted 1 to 5 are illustrated in 
Fig.~\ref{fig_schematic}.}
\label{fig_magnetization}
\end{figure}

\begin{figure}[tb]
\begin{center}
\includegraphics[width=0.48\textwidth]{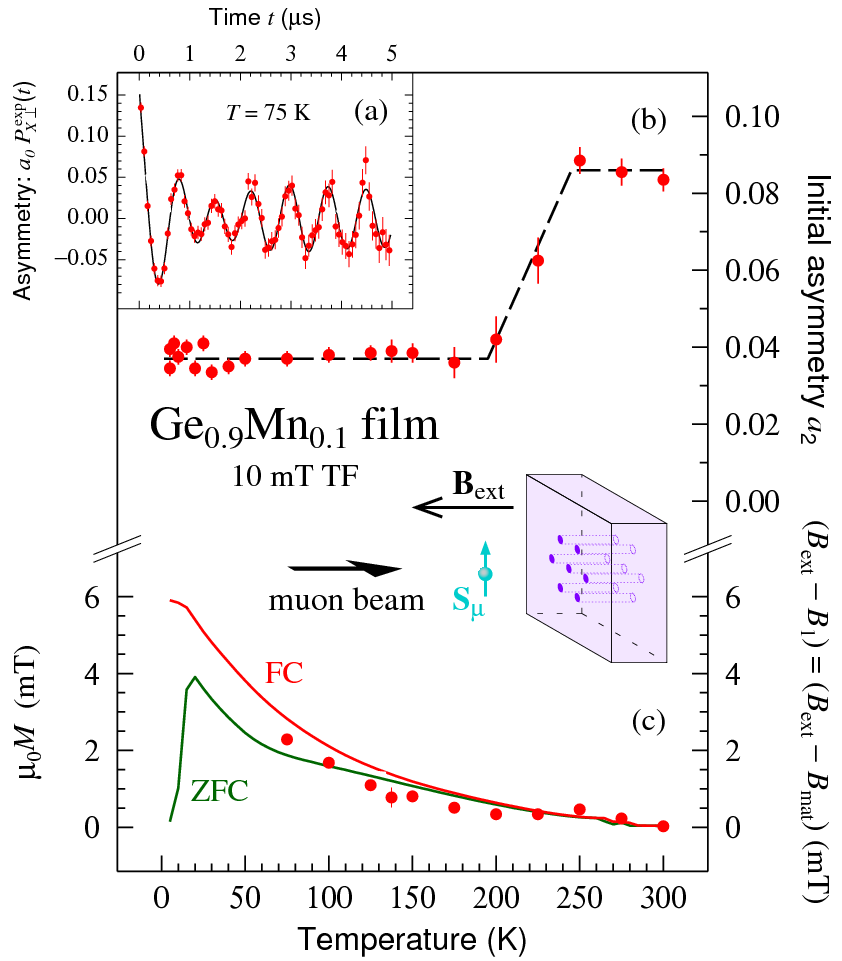}
\end{center}
\caption{(Color online)  Transverse-field $\mu$SR data with $B_{\rm ext} = 10$~mT. (a) A spectrum  
at 75~K with the result of a fit to Eq.~\ref{Fit_1}. (b) Thermal dependence of the asymmetry  $a_2$, and (c) field difference 
$(B_{\rm ext} - B_1) = (B_{\rm ext} - B_{\rm mat})$ (bullets) compared to the field $\mu_0 M$, 
where $M$ is the bulk magnetization measured either with a field or a zero-field cooling 
procedure (full lines). The TF-$\mu$SR geometry is illustrated with the pictogram. The dashed line in panel 
(b) is a guide to the eyes.
}
\label{fig_muSR_TF}
\end{figure}

We now discuss our magnetization measurements. With a Mn content less than 0.05\%, any 
magnetic order within the Ge matrix can be ruled out. From Fig.~\ref{fig_magnetization}a 
two magnetic transitions are inferred: a first one close to room temperature --- to be better defined below by 
transverse-field (TF)-$\mu$SR --- at which the nanocolumn cores order, and a second one at $\approx 70$~K 
corresponding to the onset of magnetization in the shell, probably of the spin-glass type owing to the large Mn content change. 
\begin{figure}[tb]
\begin{center}
\includegraphics[width=0.45\textwidth]{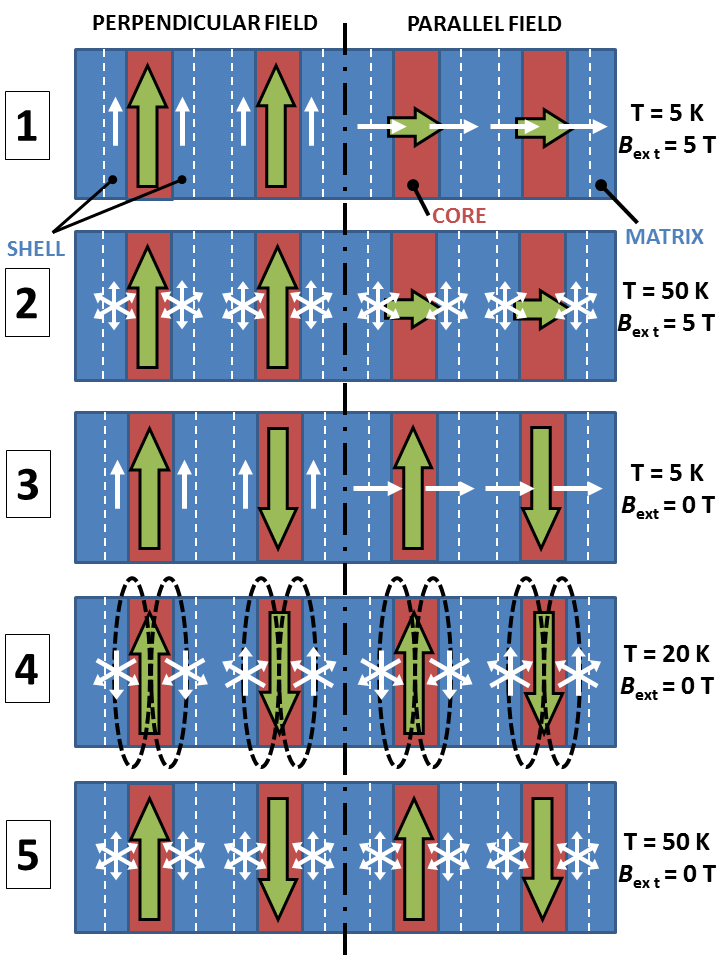}
\end{center}
\caption{(Color online) Schematic magnetic configurations for the cores and shells for fields and temperatures defined as labels in Fig.~\ref{fig_magnetization}. The ${\bf B}_{\rm ext}$ direction is relative to the film substrate. \textbf{1}: The cores and shells are both ferromagnetically (FM) aligned parallel to ${\bf B}_{\rm ext}$: $M^{\rm core}=M^{\rm core}_{s} \approx 30$~kA/m and $M^{\rm shell}=M^{\rm shell}_{s}\approx 13$~kA/m where the $s$ subscript denotes the saturation magnetization. \textbf{2}: The cores remain FM aligned while the shells become paramagnetic (PM): $M^{\rm core}=M^{\rm core}_{s}$ and $M^{\rm shell}=0$. \textbf{3}: Remanent state at 5~K after cooling under $B_{\rm ext} = 5$~T. The shell magnetic moments remain in the saturated configuration ($M^{\rm shell}=M^{\rm shell}_{s}$) whereas the cores contribution plummets because of their basically antiparallel arrangement. Still, when ${\bf B}_{\rm ext}$ is perpendicular to the film, some disorder in the columns position yields a small remanent magnetization of about 1.5~kA/m, i.e.\  $M^{\rm core}\approx M^{\rm core}_{s}/20$ whereas $M^{\rm core}\approx 0$ for the parallel setup. \textbf{4}: The stray field of a core polarizes the PM moments in the shell antiparallel to the core moment: $M^{\rm core}\approx M^{\rm core}_{s}/20$ and $M^{\rm shell} \propto -M^{\rm core}$ for the perpendicular configuration and $M^{\rm core}=M^{\rm shell}\approx 0$ in the parallel configuration. \textbf{5}: The cores remain in a remanent state while the shell moments are no longer polarized by the stray fields: $M^{\rm shell}=0$ while $M^{\rm core}\approx M^{\rm core}_{s}/20$ ($M^{\rm core}\approx 0$) in the perpendicular 
(parallel) configuration.}
\label{fig_schematic}
\end{figure}
The respective magnetizations yield the magnetic moment per Mn in both phases: 
$\mu_{\rm core}$ = 1.8 $\mu_{\rm B}$/Mn and an average $\mu_{\rm shell}$ = 0.6\,$\mu_{\rm B}$/Mn. 
These values correspond to 827~kA/m and 124~kA/m saturation magnetizations when normalized to the core and shell volumes. The magnetic configurations denoted as 1 to 5 in 
Fig.~\ref{fig_magnetization} are depicted in Fig.~\ref{fig_schematic}. Considering their nanometer size, the 
nanocolumns behave as superparamagnetic nanostructures in SQUID measurements with a maximum blocking 
temperature $T_{\rm B}^{\rm SQUID}$ = 150~K given by the vanishing remanence in 
Fig.~\ref{fig_magnetization}c and the bifurcation between zero field cooling (ZFC) and 
field cooling (FC) curves in Fig.~\ref{fig_muSR_TF}c. 

We now analyze our remanence magnetic data in Figs.~\ref{fig_magnetization}c and d. 
Because $a^{\rm lp}_{\rm nc}$ is very short an antiferromagnetic order 
of the nanocolumns should occur through the nanocolumn dipolar coupling \cite{Jamet06}. We thus expect 
no magnetic remanence at low temperature from the nanocolumns cores. In fact, at 5~K we observe quite a 
large remanent magnetic signal: almost 35\% of the saturation magnetic moment. Moreover, it is nearly 
isotropic and almost disappears at 20~K; see Fig.~\ref{fig_magnetization}d. This is the position of the 
ZFC peak in Fig.~\ref{fig_muSR_TF}. Attributing the remanence to the shells, all the results are consistent with a parallel moment alignment in the shell below a critical temperature of 
20~K. With this assignment we implicitely assume that there is no exchange magnetic coupling between the cores and shells. The remanent states denoted as 3, 4, and 5 in Figs.~\ref{fig_magnetization}c and d are explicitely 
shown in Fig.~\ref{fig_schematic}. For ${\bf B}_{\rm ext}$ perpendicular to the film substrate, a finite 
remanent signal is detected above 20~K. We attribute it to disorder within the nanocolumn array which inhibits a perfect cancelation of the core moments. The only coupling between the cores and shells is observed in configuration 4 where the paramagnetic Mn atoms in a given shell are polarized along the stray field from its core 
leading to a dip in the remanence curve of Fig.~\ref{fig_magnetization}d.

$\mu$SR spectroscopy was used to probe the nanocolumn magnetic state at the atomic scale. 
In Fig.~\ref{fig_muSR_TF}a
is shown a typical weak TF-$\mu$SR spectrum $a_0 P^{\rm exp}_{X\perp}(t)$ recorded after a FC procedure. 
$P_{X\perp}^{\rm exp}(t)$ measures the evolution of the muon 
polarization in a plane perpendicular to ${\bf B}_{\rm ext}$ \cite{Yaouanc11}. As explained in Ref.~\cite{SM}, for 
$ 40 \leq T \leq 300$~K 
\begin{eqnarray}
a_0 P^{\rm exp}_{X \perp}(t) & = & a_1
\exp[-(\lambda_{X \perp} t)^\beta]\cos(\gamma_\mu B_1 t + \varphi)\cr
& & +\, a_2  \cos(\gamma_\mu B_{\rm ext}  t + \varphi),
\label{Fit_1}
\end{eqnarray}
where $\varphi$ is a phase constant, $a_1 = 0.13$, and $B_1$ and $a_2$ are free 
parameters. The stretched exponential accounts for the magnetic inhomogeneity and dynamics of the first 
component. The exponent $\beta$ has been set in consistency with zero-field (ZF) data \cite{SM}. 
The relaxation rate $\lambda_{X\perp}$ is so large below 40~K that it cannot be determined with 
confidence. Still the second component can be analyzed after dropping the initial channels and setting 
$a_1 =0 $ in Eq. \ref{Fit_1}. The temperature dependences of $a_2$ and the field difference 
$(B_{\rm ext} - B_1)$ are displayed in Fig.~\ref{fig_muSR_TF}.

Let us discuss the first component parameters. The matrix between the columns occupying a volume bigger than that of the columns, it is natural to set $a_1 = a_{\rm mat}$, 
where $a_{\rm mat}$ is the amplitude for the muons implanted in the matrix. As a consequence,  
$B_1$ = $B_{\rm mat}$. From Fig.~\ref{fig_muSR_TF}c the field difference 
$B_{\rm mat} - B_{\rm ext}$ is on the order of $-\mu_0M$. 
The sign of this field can be understood as follows. In the absence of 
Mn atoms in the matrix, it is essentially the dipolar field produced by the nanocolumns and is 
therefore antiparallel to the column moments. We speculate the slight reduction in absolute value 
of the measured field compared to $-\mu_0M$ below 200~K (see Fig.~\ref{fig_muSR_TF}c) to arise from Mn 
moments in the shell which are slightly polarized antiparallel to the cores.

We now deal with $a_2(T)$. As the film is cooled below 250~K, $a_2$ starts to decrease: the nanocolumns 
enter a magnetic ordered state. The transition is completed at $\approx 200$~K at which $a_2(T)$ 
reaches a plateau. Therefore the transition width is $\approx 50$~K with an average 
ordering temperature $T_c^{\rm core}$ = 225~K \footnote{One could conceive the transition from the paramagnetic to superparamagnetic state of the column cores to take place above room temperature and that the blocking temperature corresponding to the TF-$\mu$SR technique is around 225~K where $a_2(T)$ drops --- while it is around $T_{\rm B}^{\rm SQUID}$ for the the SQUID measurements. Recalling that the characteristic time scale of TF-$\mu$SR is $(\gamma_\mu B_{\rm ext})^{-1}\simeq $ 1~$\mu$s in our case, this interpretation of the data would lead to an unrealistic short attempt time and therefore cannot hold.}. 
In a weak TF-$\mu$SR experiment the muons probing a magnetically ordered region --- here the columns cores --- and its vicinity are rapidly depolarized \cite{*[{See e.g.\ }] [{}] Luke94} and do not contribute to the signal. Therefore, we attribute the second component below 200~K to muons implanted in the sample surroundings, i.e.\ the so-called background with $a_{\rm bg} \simeq 0.035$, a reasonable value. Above $T_c^{\rm core}$, the amplitude $a_{\rm c} = a_2 -a_{\rm bg}\simeq 0.05$, corresponds to muons in the paramagnetic cores and their surroundings \footnote{The film volume fraction corresponding to the column cores and the region within a distance $r$ from them is $\pi (d_{\rm c}/2+r)^2/[\sqrt{3}(a^{\rm lp}_{\rm nc})^2/2]$ which is equal to $a_{\rm c}/(a_{\rm c}+a_{\rm mat})$ provided that $r$ = 1.8~nm. Such a $r$ value is in the accepted range; see e.g.\ \cite{Niedermayer98}}.

The spin dynamics in our film has been investigated by ZF-$\mu$SR measurements. A 
selection of $a_0 P^{\rm exp}_{Z\,\parallel,\perp}(t)$ spectra with the initial muon 
polarization ${\bf S}_\mu$ either parallel or perpendicular to the film substrate
\begin{figure}[tb]
\begin{center}
\includegraphics[width=0.45\textwidth]{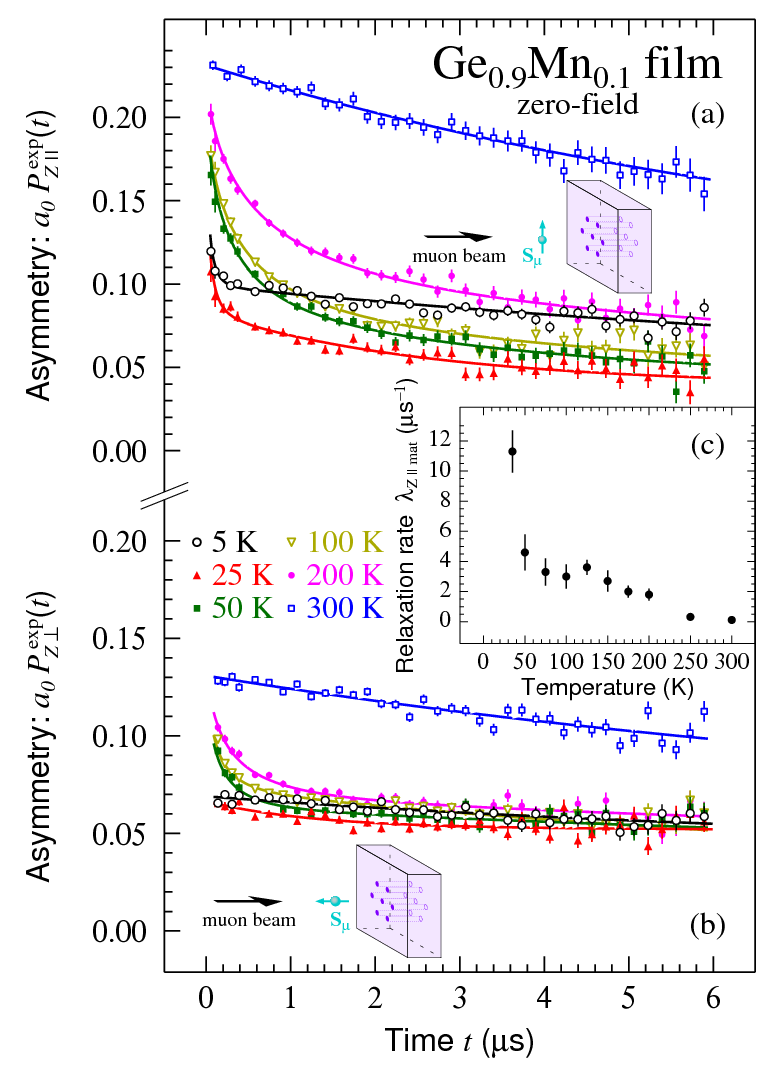}
\end{center}
\caption{(Color online) A selection of ZF-$\mu$SR spectra for our Ge$_{0.9}$Mn$_{0.1}$ film 
with ${\bf S}_\mu$ either parallel (panel a) or perpendicular (panel b) to the film substrate. 
The solid lines arise from fits. The thermal dependence of the $\mu$SR spin-lattice 
relaxation rate $\lambda_{Z\,\parallel\,{\rm mat}}$ is displayed in panel (c).}
\label{fig_muSR_ZF}
\end{figure}
is shown in Fig.~\ref{fig_muSR_ZF}. $P^{\rm exp}_{Z\,\parallel,\perp}(t)$ denotes the evolution of the 
projection of the muon polarization along ${\bf S}_\mu$. 
We note three qualitative characteristics. 
(i) Given the uniaxial character of the nanocolumns, 
the similitude of the spectra for the two geometries is striking. 
Since we probe the dynamics of the Mn spins, we conclude that we are dealing with 
isotropic spins, consistent with a Mn$^{2+}$ electronic state.
(ii) The transition from the paramagnetic to the 
superparamagnetic states is fingerprinted by comparing the spectra at 300 and 200~K. 
(iii) While the spectral shape monotically evolves as the 
sample is cooled down to 25~K, the relaxation decreases between 25 and 5~K.
A quantitative analysis is given in \cite{SM}. The
spin-lattice relaxation rate in the matrix $\lambda_{Z\,\parallel\,{\rm mat}}$ is proportional to the fluctuation time $\tau$ of the Mn spins in the shells. From the time window of the $\mu$SR technique \cite{Yaouanc11}, an upper bound for $\tau$ is in the microsecond range. A sharp increase of $\tau$ is observed below 50~K (Fig.~\ref{fig_muSR_ZF}). This is consistent with the remanent 
magnetization which shows the Mn spins to relax on the SQUID time scale, i.e.\ on the order of a  
minute at $\approx$ 20~K and to be completely frozen at 5~K. 

In conclusion, our Ge$_{0.9}$Mn$_{0.1}$ film of self-organized nanocolumns is 
a complex object. The columns comprise a core of diameter 2~nm with a nearly 70\% Mn atomic 
concentration, and a shell of thickness 1~nm in which the Mn concentration decreases to 0. The Mn 
concentration in the matrix between the columns is below the detection limit of 0.05\%. A transition 
from a paramagnetic to a superparamagnetic state is observed for the cores around 225 K. Below 150~K 
they start to freeze on the SQUID time scale. No magnetic order is found for the Mn moments of the 
shells. Their spin dynamics is rather isotropic, consistently with a Mn$^{2+}$ state. The shell magnetic moments start to slow down on the microsecond time scale below 50~K and become frozen at 5~K . There is no magnetic coupling between a core and its shell, except for a weak dipolar coupling. Finally we infer that the separation in three distinct structural and magnetic regions observed here is generic to semiconductors doped with a non miscible transition element.

Part of this work has been performed at the Swiss Muon Source (S$\mu$S),
Paul Scherrer Institute, Villigen, Switzerland.  We thank J.-F. Jacquot for his help
during the SQUID measurements.

\bibliographystyle{proposal}
\bibliography{spintronics,GeMn_letter_SM}

\end{document}


\title{Supplemental Material: \\
Core-shell nanostructure in a Ge$_{0.9}$Mn$_{0.1}$ film from structural and magnetic measurements}

\author{P.~Dalmas de R\'eotier}
\affiliation{Univ. Grenoble Alpes, INAC-SPSMS, F-38000 Grenoble, France}
\affiliation{CEA, INAC-SPSMS, F-38000 Grenoble, France}
\author{E.~Prestat}
\affiliation{Univ. Grenoble Alpes, INAC-SP2M, F-38000 Grenoble, France}
\affiliation{CEA, INAC-SP2M, F-38000 Grenoble, France}
\author{P.~Bayle-Guillemaud}
\affiliation{Univ. Grenoble Alpes, INAC-SP2M, F-38000 Grenoble, France}
\affiliation{CEA, INAC-SP2M, F-38000 Grenoble, France}
\author{A.~Barski}
\affiliation{Univ. Grenoble Alpes, INAC-SP2M, F-38000 Grenoble, France}
\affiliation{CEA, INAC-SP2M, F-38000 Grenoble, France}
\author{A.~Marty}
\affiliation{Univ. Grenoble Alpes, INAC-SP2M, F-38000 Grenoble, France}
\affiliation{CEA, INAC-SP2M, F-38000 Grenoble, France}
\author{M.~Jamet}
\affiliation{Univ. Grenoble Alpes, INAC-SP2M, F-38000 Grenoble, France}
\affiliation{CEA, INAC-SP2M, F-38000 Grenoble, France}
\author{A.~Suter}
\affiliation{Laboratory for Muon-Spin Spectroscopy, Paul Scherrer Institute, 
CH-5232 Villigen-PSI, Switzerland}
\author{T.~Prokscha}
\affiliation{Laboratory for Muon-Spin Spectroscopy, Paul Scherrer Institute, 
CH-5232 Villigen-PSI, Switzerland}
\author{Z.~Salman}
\affiliation{Laboratory for Muon-Spin Spectroscopy, Paul Scherrer Institute, 
CH-5232 Villigen-PSI, Switzerland}
\author{E.~Morenzoni}
\affiliation{Laboratory for Muon-Spin Spectroscopy, Paul Scherrer Institute, 
CH-5232 Villigen-PSI, Switzerland}
\author{A. Yaouanc}
\affiliation{Univ. Grenoble Alpes, INAC-SPSMS, F-38000 Grenoble, France}
\affiliation{CEA, INAC-SPSMS, F-38000 Grenoble, France}

\date{\today}

\begin{abstract}
We provide information on the transmission electron microscopy (TEM) and electron energy loss 
spectrometry (EELS) techniques we have used, and show additional TEM images, maps and profiles 
obtained from EELS for our Ge$_{0.9}$Mn$_{0.1}$ film. Complementary information on the muon spin relaxation 
measurements and their analysis is given.
\end{abstract}

\maketitle

\section{Transmission electron microscopy and electron energy loss spectrometry characterization}

\begin{figure}[h!]
\begin{center}
\includegraphics[width=0.475\textwidth]{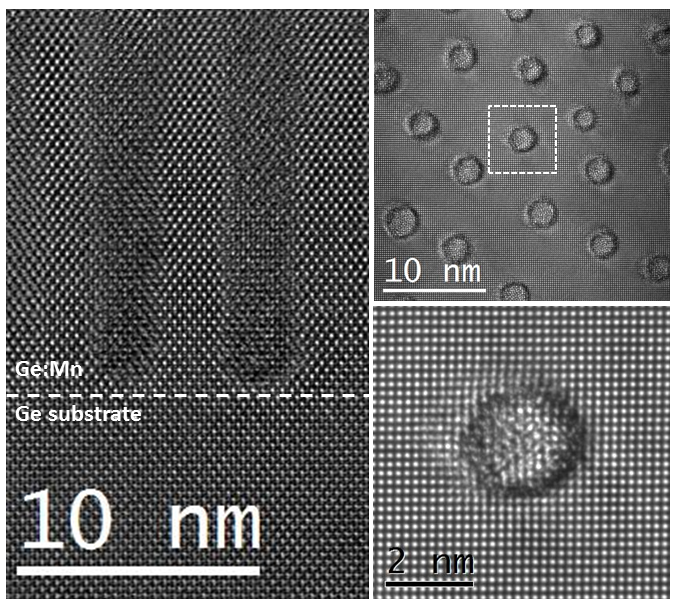}
\end{center}
\caption{Low and high resolution TEM images of our Ge$_{0.9}$Mn$_{0.1}$ film.}
\label{supplementary_fig1}
\end{figure}

Transmission electron microscopy (TEM) measurement were performed using a monochromated and 
double-corrected 80-300~kV FEI Titan$^{3}$ Ultimate microscope working at 200~kV. Taking advantage of the 
image-side aberration corrector, high resolution TEM (HRTEM) images were acquired close to Gaussian 
focus, i.e.\ within a few nanometers away to avoid delocalization effect. To increase the signal-to-noise 
ratio of the HRTEM measurement, series of about 10 images have been acquired from the same area. After 
alignement of the image series using a cross-correlation based algorithm, we averaged the images over the 
series, to obtain an HRTEM image with very good signal-to-noise ratio. TEM images of a Ge$_{0.9}$Mn$_{0.1}$ 
film grown at 100$^{\circ}$C are shown in Fig.~\ref{supplementary_fig1}

Electron energy loss spectroscopy (EELS) acquisitions were achieved using a Gatan imaging filter (GIF) Quantum
in dual-EELS mode with an energy dispersion of 1~eV, allowing to acquire a full 
range electron loss from 0 to 1400~eV. The convergence and collection angles were 20 and 98~mrad, 
respectively. Typical pixel acquisition time was 20-50~ms and the typical map size was about $200\times200$ 
pixels. The pixel size was set to 0.3~nm. The noise of the EELS signal was removed using a principal component analysis (PCA) algorithm 
implemented in the open-source software Hyperpsy.\cite{hyperspy} The relevant number of components for the 
reconstruction of the spectra was chosen by careful inspection of the factors and loadings 
of the first 10 components. For each edge (Mn~L$_{2,3}$ at 640~eV and Ge~L$_{2,3}$ at 1217~eV), the background 
was removed by subtracting a power law fitted in the region preceding the edge with a width of 40-80 eV 
(depending on the edge). Element maps were obtained by integrating over 60~eV after the edge for Mn and Ge.

Theoretically, the signal of ionization edge $I_{k}$ can be
quantified using standardless method, following the relation:
\begin{equation}
I_{k}=n_{\rm{den}}I\sigma_{k}\label{eq:count_edge}
\end{equation}
where $n_{\rm{den}}$ is the areal density
(atom/nm\textsuperscript{2}) of a given element, $I$ the total
integrated number of counts in the spectrum and $\sigma_{k}$ the
cross section of ionization of an electron in the corresponding electron shell.
Taking advantage of the high dynamical range of the spectra provided by the acquisition in dual-EELS mode, 
we can measure the total integrated number of counts $I$ (Eq.~\ref{eq:count_edge}) in the spectrum --- from 
the intense zero-loss peak to the Ge and Mn edges. 

In practice, the integration is performed over a finite energy range $\Delta$ and the collection angle 
$\beta$ is limited by the entrance aperture of the spectrometer. These experimental parameters are taken 
into account in the calculation of the partial cross-sections, which depend on $\Delta$, on the collection 
angle $\beta$ and on the energy loss. For the Ge and Mn quantification we calculated parametrized partial 
Hartree-Slater cross sections \cite{Egerton11} using the Gatan Digital Micrograph software.

By further measuring the specimen thickness, we can calculate the absolute volumic density 
(atom/nm\textsuperscript{3}) of each species.

\begin{figure}[tb]
\begin{center}
\includegraphics[width=0.475\textwidth]{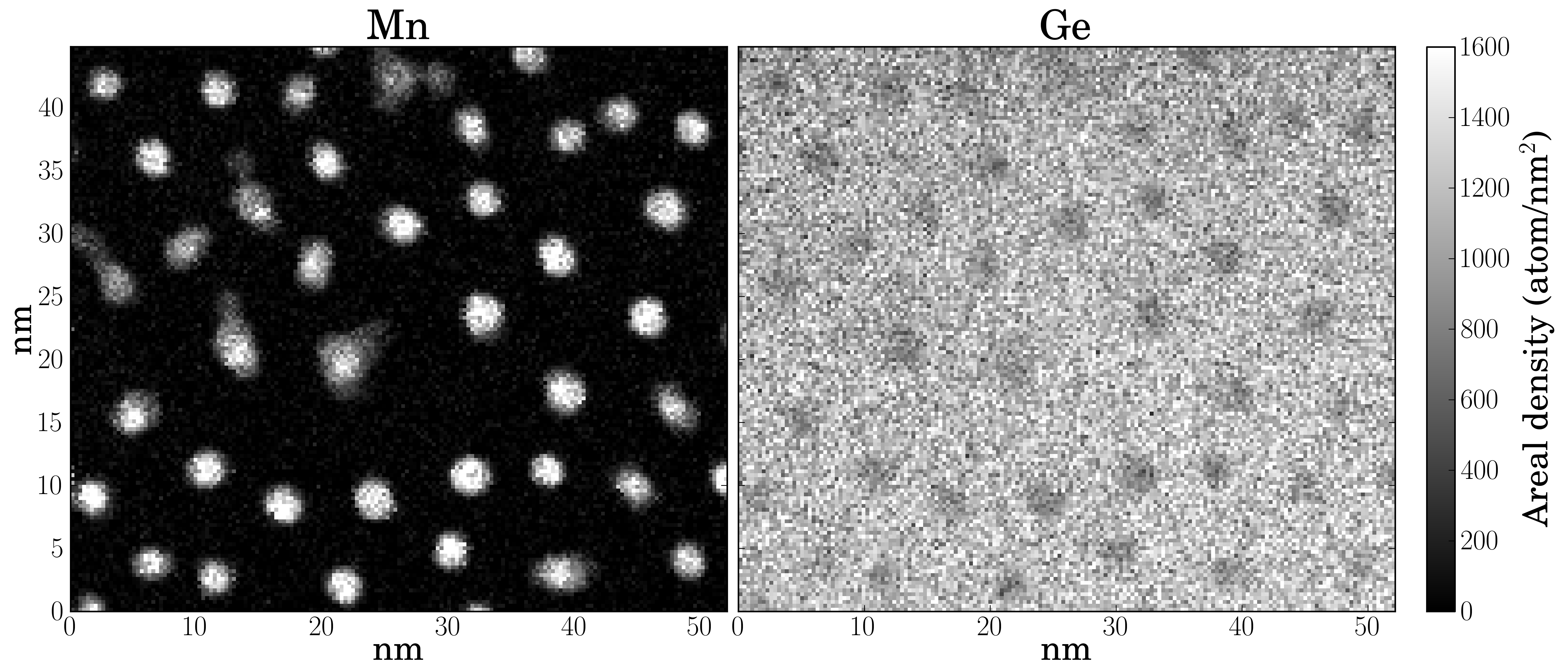}
\end{center}
\caption{Mn and Ge areal density maps obtained by EELS measurements.}
\label{supplementary_fig2}
\end{figure}

\begin{figure}[tb]
\begin{center}
\includegraphics[width=0.475\textwidth]{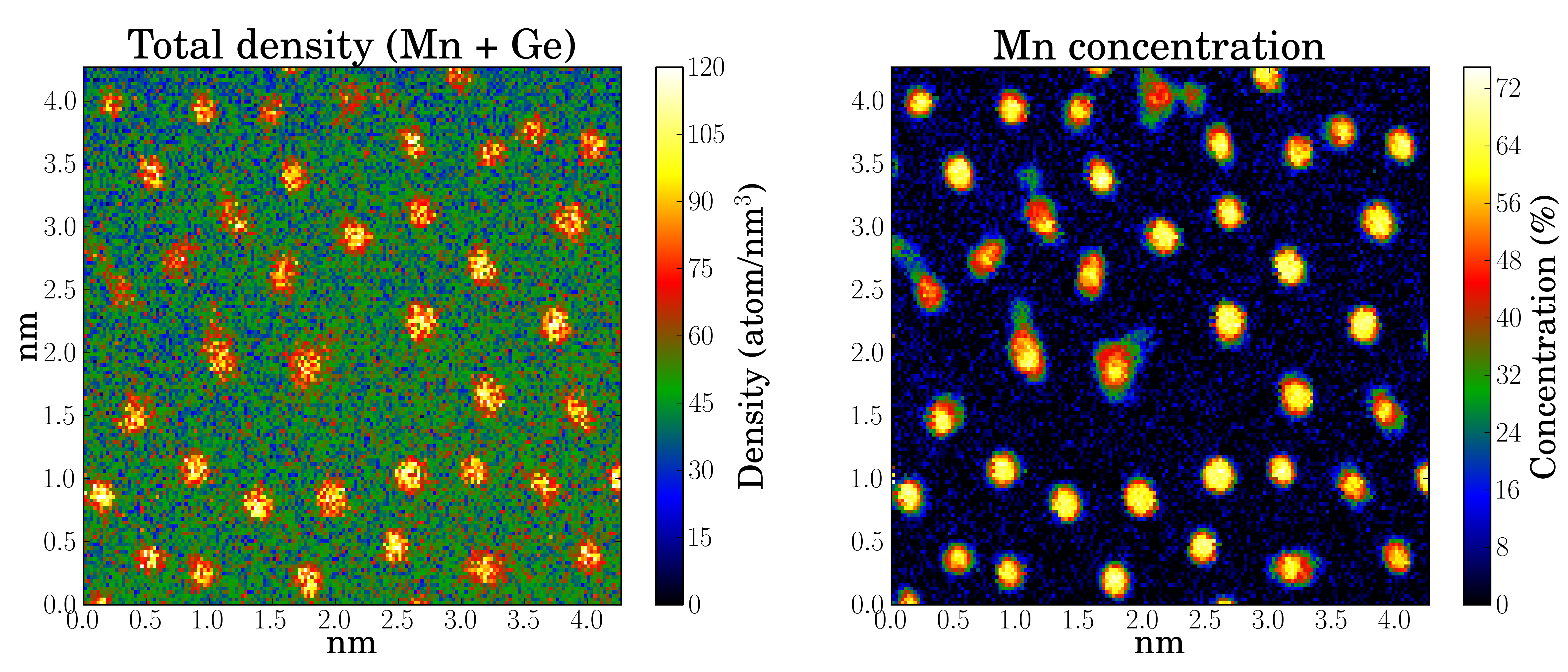}
\end{center}
\caption{Total density and Mn relative concentration maps obtained by EELS measurements.}
\label{supplementary_fig3}
\end{figure}

\begin{figure}[tb]
\vspace{4mm}
\begin{center}
\includegraphics[width=0.475\textwidth]{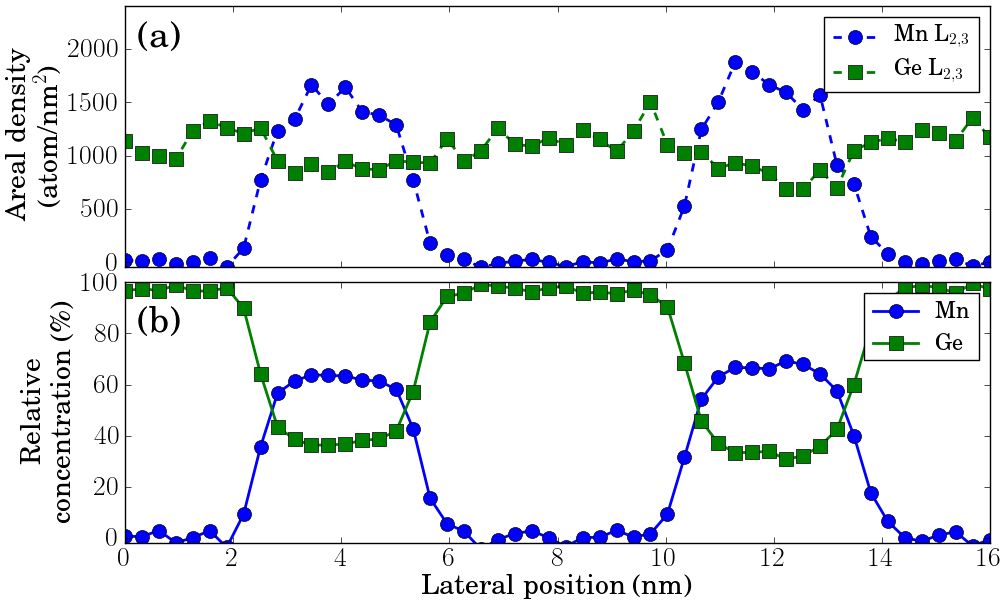}
\end{center}
\caption{(a) Areal density and (b) relative concentration profiles measured across two nanocolumns.}
\label{supplementary_fig4}
\end{figure}

To check the reliability of our analysis, we performed two cross checks:
\begin{itemize}
\item The experimental value for the Ge volumic density in the diamond Ge pure matrix is in good 
agreement --- within 10~\% --- with the corresponding theoretical value (44~atoms/nm\textsuperscript{3}).
\item The Mn concentration averaged over the whole concentration map is in good agreement --- within 
2\% --- with the Mn nominal concentration of the Ge(Mn) film measured with Rutherford backscattering 
spectrometry (RBS).
\end{itemize}
Figure~\ref{supplementary_fig2} 
shows the Mn and Ge areal density maps obtained by the method explained previously.
The total density (Mn + Ge density) is calculated by dividing the total areal density (Mn + Ge areal density) 
by the specimen thickness. Figure~\ref{supplementary_fig3} 
shows the total density and the Mn relative concentration maps. Profiles measured on two nanocolumns are 
plotted in Fig.~\ref{supplementary_fig4}.


\section{Muon spin measurements}

In Sec.~\ref{Muon_profile} we display the muon implantation profile computed for muons with
7.5~keV kinetic energy. 
Section~\ref{Muon_TF}
provides additional information on the TF-$\mu$SR measurements. We finally describe the detailed
analysis of the ZF-$\mu$SR data in Sec.~\ref{Muon_ZF}.

\subsection{Profile of muon implantation}
\label{Muon_profile}

\begin{figure}[tb]
\vspace{4mm}
\begin{center}
\includegraphics[width=0.475\textwidth]{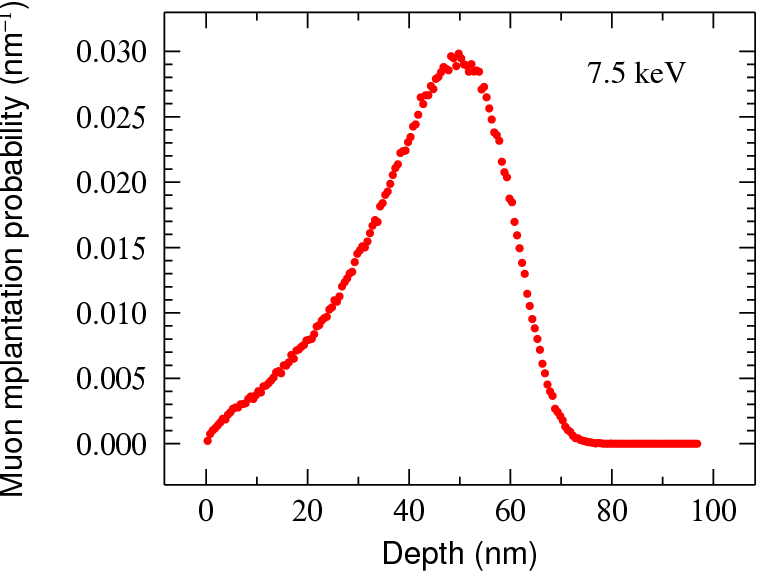}
\end{center}
\caption{Distribution of the depth implantation profile for muons of 7.5~keV kinetic energy. The 
simulation has been performed using the Monte Carlo code TRIM.SP (Refs.~\onlinecite{Eckstein91,Morenzoni04,Morenzoni01}) code for a GeMn film with a Ge:Mn atomic ratio of 9:1.}
\label{muon_implantation}
\end{figure}

Muons of 7.5~keV kinetic energy were used for the $\mu$SR measurements reported 
in the main text. From the implantation depth profile shown in Fig.~\ref{muon_implantation}, 
we can conclude that all the muons are implanted within the film thickness of 80~nm.

\subsection{Transverse field $\mu$SR}
\label{Muon_TF}

The TF-$\mu$SR spectrum displayed in Fig.4a of the main text exhibits two beating oscillating components. 
A first analysis  for $ 100 \leq T \leq 200$~K indicates the amplitude of the more rapidly damped 
component to be temperature independent. Below 100~K the damping of this first component becomes quite 
large. As a consequence its initial amplitude, damping, precession frequency and initial phase are 
determined with limited precision. Fixing the initial amplitude allows a better accuracy on the other three 
parameters down to 40~K. To a good approximation, the precession frequency of the second 
component is constant in the whole temperature range with an average 
$\nu_2 = 1.353$~MHz. This corresponds to $B_{\rm ext}$, i.e.\ $\nu_2 = \nu_{\rm ext}$  = 
$\gamma_\mu B_{\rm ext}/(2 \pi)$. Here $\gamma_\mu$ = 851.6~Mrad\,s$^{-1}$\,T$^{-1}$ is 
the muon gyromagnetic ratio. This discussion justifies the formula of Eq.~1 in the main text which is used
to analyze the TF-$\mu$SR spectra.

\subsection{Zero field $\mu$SR}
\label{Muon_ZF}

From the two spectra recorded at 300~K, see Fig.~5 of the main text, we measure $a_0 \simeq 0.23$ (0.13) when 
${\bf S}_\mu$ is parallel (perpendicular) to the film substrate. The smaller $a_0$ value for the perpendicular 
geometry follows from the geometrical setup of the spectrometer. 

A qualitative discussion of the ZF-spectra is given in the main text. Here we quantitatively describe
them. The measured asymmetry can be modelled  consistently with the 
TF-$\mu$SR model \footnote{Given the similitude of the data recorded for the initial muon polarization perpendicular and parallel to the film and since $a_0$ is much larger for the latter geometry, we concentrate our discussion on it.}
\begin{eqnarray}
a_0 P^{\rm exp}_{Z\,\parallel}(t)  =  
\alpha &\, & \left\{ a_{\rm mat} \exp[-(\lambda_{Z\,\parallel\,{\rm mat}} t)^\beta]\right. \cr
 & &\ \  +  \left. a_{\rm c} \exp(-\lambda_{Z\,\parallel\,{\rm c}} t) 
+a_{\rm bg} \right\},
\label{Fit_2}
\end{eqnarray}
where $\lambda_{Z\,\parallel\,{\rm mat}}$ and $\lambda_{Z\,\parallel\,{\rm c}}$ are spin-lattice 
relaxation rates. The $\alpha = 1.05$ geometrical parameter is introduced to account for the slight 
difference in the signal amplitude between the TF and ZF geometries. Naturally we assign the first 
(second) term in Eq.~\ref{Fit_2} to muons stopped in the matrix (cores and their vicinity) while the 
last one concerns muons which have missed the sample. We find values for $a_{\rm mat}$, $a_{\rm c}$, 
and $a_{\rm bg}$ in line with the TF results. Fitting Eq.~\ref{Fit_2} to the data, we obtain 
$\lambda_{Z\,\parallel\,{\rm mat}}(T)$ (see Fig.~5c of the main text), $\lambda_{Z\,\perp\,{\rm c}}(T)$ 
and an exponent $\beta$ equal to 0.6 between 200 and 50~K with a tendency to increase at lower 
temperature. At room temperature, $\beta = 1$ provides a good fit to the data.

Recalling the results obtained in TF measurements, and the structure of the film, one can deduce 
that the distance from muons concerned by the first component in Eq.~\ref{Fit_2} to the nearest Mn 
ions spans between 0.8 and 3~nm. This is a situation reminiscent of the dilute canonical spin glasses. 
In such systems, the polarization function has been derived in the case of fast fluctuating moments
to be $P_{Z}(t) = \exp[-\sqrt{\lambda_{Z} t}]$,\cite{Uemura85,Yaouanc11} a function which is 
close to our observation. The relaxation rate $\lambda_{Z}$ is related to the magnetic fluctuation 
time $\tau$ of the Mn spins in the shell through the formula 
$\lambda_{Z}$ = $4\gamma_\mu^2\Delta^2_{\rm L}\tau$, where $\Delta_{\rm L}$ is the half-width at 
half-maximum of the component field Lorentzian distribution.\cite{Walstedt74} 

Finally a few ZF spectra have been recorded after prior cooling of the film 
with $B_{\rm ext} = 150$~mT perpendicular to the film substrate. Within the data error bars no 
difference is found with the spectra measured after the usual ZFC procedure.

\bibliography{spintronics}